%% file: 0_main.tex
\documentclass[sigconf, nonacm]{acmart}
\usepackage{multirow}
\usepackage{multirow}
\usepackage{colortbl}
\usepackage{bm}
\usepackage{enumitem}       
\usepackage{bbding}       
\usepackage{microtype}
\usepackage{multicol}
\usepackage[section]{placeins}
\usepackage{graphicx}
\usepackage{stfloats}
\usepackage{enumerate}

\newcommand{\ie}{\emph{i.e., }}
\newcommand{\eg}{\emph{e.g., }}

\newcommand{\etc}{\emph{etc.}}
\newcommand{\wrt}{\emph{w.r.t. }}

\AtBeginDocument{%
  \providecommand\BibTeX{{%
    \normalfont B\kern-0.5em{\scshape i\kern-0.25em b}\kern-0.8em\TeX}}}

\begin{document}
\begin{sloppypar}

\title{Headache to Overstock? Promoting Long-tail Items through Debiased Product Bundling}





\author{Shuo Xu$^*$}
\affiliation{%
  \institution{School of Software, Shandong University} 
  \country{China}
}
\email{shuo.xu@mail.sdu.edu.cn}

\author{Haokai Ma$^*$}
\affiliation{%
  \institution{School of Software, Shandong University} 
  \country{China}
}
\email{mahaokai@mail.sdu.edu.cn}

\author{Yunshan Ma}
\affiliation{%
  \institution{National University of Singapore} 
  \country{Singapore}
}
\email{yunshan.ma@u.nus.edu}

\author{Xiaohao Liu}
\affiliation{%
  \institution{National University of Singapore} 
  \country{Singapore}
}
\email{xiaohao.liu@u.nus.edu}

\author{Lei Meng$^\dagger$}
\affiliation{%
  \institution{Shandong Research Institute of Industrial Technology; School of Software, Shandong University} 
  \country{China}
}
\email{lmeng@sdu.edu.cn}

\author{Xiangxu Meng}
\affiliation{%
  \institution{School of Software, Shandong University} 
  \country{China}
}
\email{mxx@sdu.edu.cn}

\author{Tat-Seng Chua}
\affiliation{%
  \institution{National University of Singapore} 
  \country{Singapore}
}
\email{dcscts@nus.edu.sg}
\thanks{$^*$ indicates equal contribution.}
\thanks{$^\dagger$ indicates corresponding author.}

\setcopyright{none}

\begin{abstract}
Product bundling aims to organize a set of thematically related items into a combined bundle for shipment facilitation and item promotion. To increase the exposure of fresh or overstocked products, sellers typically bundle these items with popular products for inventory clearance. This specific task can be formulated as a long-tail product bundling scenario, which leverages the user-item interactions to define the popularity of each item. The inherent popularity bias in the pre-extracted user feedback features and the insufficient utilization of other popularity-independent knowledge may force the conventional bundling methods to find more popular items, thereby struggling with this long-tail bundling scenario. 
Through intuitive and empirical analysis, we navigate the core solution for this challenge, which is maximally mining the popularity-free features and effectively incorporating them into the bundling process.
To achieve this, we propose a \textbf{D}istilled Modal\textbf{i}ty-Oriented Knowl\textbf{e}dge \textbf{T}ransfer framework (DieT) to effectively counter the popularity bias misintroduced by the user feedback features and adhere to the original intent behind the real-world bundling behaviors.
Specifically, DieT first proposes the Popularity-free Collaborative Distribution Modeling module (PCD) to capture the popularity-independent information from the bundle-item view, which is proven most effective in the long-tail bundling scenario to enable the directional information transfer. With the tailored Unbiased Bundle-aware Knowledge Transferring module (UBT), DieT can highlight the significance of popularity-free features while mitigating the negative effects of user feedback features in the long-tail scenario via the knowledge distillation paradigm. 
Extensive experiments on two real-world datasets demonstrate the superiority of DieT over a list of SOTA methods in the long-tail bundling scenario. 
The code will be released upon acceptance.
\end{abstract}


\copyrightyear{}
\acmYear{}
\acmDOI{}
\acmConference{}
\acmISBN{ }

\maketitle

\input{1_introduction}

\input{2_related_work}
\input{3_motivation_analysis}
\input{4_methodology}

\input{5_experiments}

\input{6_conclusion}

\balance
\bibliographystyle{ACM-Reference-Format}
\bibliography{0_main}

\end{sloppypar}
\end{document}

%% file: 1_introduction.tex
\section{Introduction} \label{sec:introduction}

Product bundling aims to curate a set of items closely related to the same theme for sale as one combined bundle \cite{BGN, BGCN}, which has been widely applied in e-commerce, online entertainment, and \etc \cite{CrossCBR, multicbr}. 
Its salient purpose in these real-world online applications is to increase the exposure of overstocked products, which are often characterized as long-tail items \wrt user-item interactions.
Specifically, these long-tail (short as LT) items are purposefully bundled with certain popular (short as Pop) items~\footnote{The popularity discussed here is solely related to the interaction number between users and items, which is commonly regarded the most crucial in online serving.}, which are defined as Pop-to-LT product bundling scenarios~\footnote{Pop-to-LT: Given some popular items as query, we aim to find some long-tail items to complete the bundle (\eg attract customers with a trendy dress and pair it with a less popular yet matching handbag).}.



\begin{figure}[!t]
    \centering
    \includegraphics[width=0.95\linewidth]{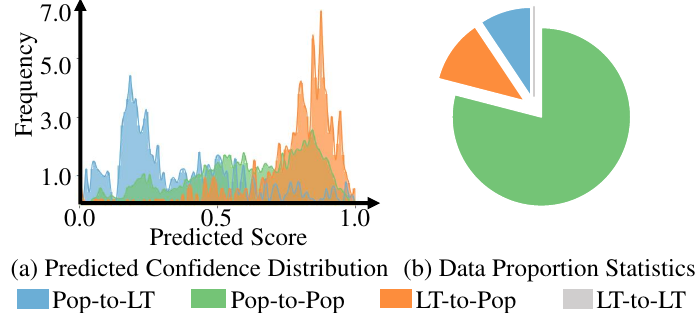}
    \caption{The predicted confidence distribution and the data proportion statistics of CLHE on four bundling settings (distinguished by colors). Notably, Pop-to-LT scenario is the most challenging with its remarkable data proportion.}
    \label{fig.motivation}
\end{figure}

Despite the significance of bundles in various online applications, automatic product bundling receives little scrutiny in academic and industrial communities. Among the bundle-related studies, only CLHE \cite{CLHE} serves as the pioneering work to explore this important but easy-to-ignore research frontier via effectively utilizing bundle-item affiliation, user-item feedback, and multimodal content knowledge.
However, mechanically integrating these three multi-view information in CLHE might yield results that deviate from the original intent of bundle construction. Moreover, the introduction of user-item feedback information in overall product bundling performance has led to unequal treatment of long-tail items and shows poor performance in the specific Pop-to-LT scenario.
Scrutinizing this problem, we conduct a systematic and empirical study across the product bundling scenarios and evaluate the impact of various features from diverse views. Specifically, we surgically break down the bundling setting into four sub-settings, \ie LT-to-Pop, Pop-to-LT, Pop-to-Pop, and LT-to-LT, regarding the popularity of the items in both query and output partial bundle. 
Then we implement CLHE \cite{CLHE} on these sub-settings to obtain its predicted confidence on the ground truth item with the given partial bundle, compile the data proportion statistics of each scenario and visualize them in Figure.~\ref{fig.motivation}, respectively.
Investigating these sub-settings, it is easily found that the Pop-to-LT setting poses the highest challenge (since it has the lowest confidence in the predicted confidence distribution of Figure.~\ref{fig.motivation}), while retaining a non-negligible portion of cases (as shown in the data distribution of Figure.~\ref{fig.motivation})~\footnote{The LT-to-LT setting has few data while the LT-to-Pop and Pop-to-Pop settings already perform pretty well.}.

Zooming into the Pop-to-LT setting, a natural question arises:
\emph{what are the key factors that cause the deteriorated performance?} To answer this, we identify two potential reasons by intuitive and empirical analysis as follows. 
First, \emph{\textbf{item-level user feedback features are not well aligned with the bundling objective in the scenario of Pop-to-LT}}.
These features are pre-extracted from collaborative filtering on the user-item view, which may force the model to prioritize more popular items
\cite{CLHE, lightgcn}. 
The indiscriminate incorporation of these features increases the ranking of popular items, hence leading to disastrous low-ranking for low-popularity items. Further surgical analysis (see Section~\ref{sec.ma_sa}) empirically demonstrates that removing the user feedback features will not hinder the performance and sometimes even improve it in the Pop-to-LT setting.
Second, \emph{\textbf{the popularity-free features, \ie bundle-item affiliation and item multimodal features, are not maximally utilized}}.
Interestingly, we find that removing bundle-item affiliation or item multimodal features results in a significant performance drop in the Pop-to-LT setting, justifying our hypothesis.
In summary, the crux to tackling the Pop-to-LT problem lies in maximally mining the popularity-free features and effectively integrating them into the bundling process.

To achieve this, we propose a novel \textbf{D}istilled Modal\textbf{i}ty-Oriented Knowl\textbf{e}dge \textbf{T}ransfer framework (DieT) to highlight the popularity-free knowledge in product bundling via teacher-student knowledge distillation, providing the antidote for long-tail items.
Inspired by the essential role of defined popularity-free knowledge in the Pop-to-LT scenario, DieT designs a Popularity-free Collaborative Distribution Modeling module (PCD) to capture the bundle-item hierarchical collaborative distribution through an intra-view self-attention mechanism.
To incorporate popularity-independent information into the Pop-to-LT scenario, DieT further proposes an Unbiased Bundle-aware Knowledge Transferring module (UBT) to enable the directional knowledge transfer from the bundle-item view to the multimodal view through the knowledge distillation paradigm. This can precisely mitigate inherent popularity bias in user behaviors and effectively promote the long-tail items.
To evaluate our framework, we conduct extensive experiments on two real-world datasets. The results show that our method significantly improves performance in the Pop-to-LT product bundling scenario while retaining high generalization and robustness in diverse settings. 
The contributions of this work can be summarized as follows:
\begin{itemize}[leftmargin=*, topsep=0.2pt,parsep=0pt]
  \item To the best of our knowledge, we are the first to identify the scenario of Pop-to-LT product bundling with extensive intuitive and empirical study for the sake of long-tail item promotion. 
  %
  \item 
  We examine the critical reasons for inferior Pop-to-LT performance as being the inadequate utilization of popularity-free features, based on which we propose a novel model-agnostic knowledge distillation-based framework DieT. 
  \item Our method achieves significant performance gain over the SOTA methods in the Pop-to-LT bundling scenario. Further analysis demonstrates its robustness and generalization capabilities.
\end{itemize}

%% file: 2_related_work.tex
\section{Related Work}
We briefly review the relevant literature for this work, including bundle recommendation and construction, knowledge distillation and popularity debiasing in recommendation.

\noindent
\textbf{Bundle Recommendation and Construction.}
Inspired by the significance of bundles in real-world applications, several recent studies have provided a summary and analysis about bundle-aware tasks from various perspectives \cite{sun2022revisiting, bundle-survey, mealrec}.
The bundle recommendation aims to provide a set of inter-correlated items rather than individual ones to users, satisfying their diverse demands in a one-stop manner. 
BRP \cite{BRP} first tackles this issue by constructing expected reward vectors and cross-dependency matrices.
BGCN \cite{BGCN} decouples user preferences into both item views and bundle views to effectively capture dual user preferences. CrossCBR \cite{CrossCBR} further emphasizes the importance of collaborative association in BGCN \cite{BGCN} using the cross-view contrastive learning paradigm to enhance the self-discrimination of representations. 
MultiCBR \cite{multicbr} adopts an "early fusion, late contrast" strategy in bundle-item affiliations, enhancing user preference modeling while minimizing extra costs.
BundleMCR \cite{bundleMCR} defines bundle recommendation as an innovative multi-round conversational procedure and leverages the RL paradigm to model it as a Markov decision process problem with multiple agents.
To further facilitate the bundle recommendation task, AICL \cite{sun2024adaptive} deduces the underlying intentions rooted in users’ behaviors within a session, employing a large language model to execute the personalized bundle generation process.


With the explosive growth of the item corpus, manually constructing large-scale bundles is quite formidable.
Thus, an automatic bundle construction strategy is indispensable and proven to be valid. 
CLHE \cite{CLHE} constructs the partial bundle with multimodal features, user feedback at the item level, and existing bundles, achieving the performance of SOTA construction.
LARP \cite{LARP} integrates multimodal and relational signals into feature representations and tackles the playlist construction task through a three-layer contrastive learning framework.
CIRP \cite{CIRP} leverages cross-item contrastive loss and individual item's image-text contrastive loss to embed modeling capabilities into a multimodal encoder, producing enhanced item representations for downstream product bundling applications.
BundleMLLM \cite{bundle-MLLM} introduces a novel multimodal large language model framework that integrates multimodal features of items and formulates the bundle construction task as a multiple-choice question with candidates as options, which could hardness the extensive external knowledge of LLMs to facilitate the bundling process.
Nevertheless, the pre-extracted item-level user feedback features emphasized in CLHE may intensify the popularity bias in the user-item view, hindering its alignment with the bundling objective of this specific Pop-to-LT scenario.

\noindent
\textbf{Knowledge Distillation in Recommendation.}
The core of knowledge distillation (KD) is that a teacher model injects hidden knowledge extracted from the data into a simpler student model, thereby enhancing the student model's performance in a specific aspect
\cite{2015distilling}.
Inspired by its success \cite{mimicking, nlp, Prod}, KD has been widely applied in recommendation \cite{Ranking_Distillation, DERRD, collaborative_distillation}. 
For instance, ALDI \cite{ALDI} designs three distilled aligning objectives to transfer behavioral information from warm items to cold ones for cold-start problems. 
SGFD \cite{SGFD} proposes response-based and feature-based distillation losses to facilitate the rich modality features transfer in multimodal recommendation.
DGMAE \cite{DGMAE} is the most relevant study to our work, which extracts the first- and higher-order associations on the user-bundle view to identify users' potential bundling preferences in bundle recommendation tasks.
To the best of our knowledge, DieT is the first KD framework for product bundling tasks.

\noindent
\textbf{Popularity Debiasing in Recommendation.}
The user behavioral data originates from observed data rather than the ideal experimental data, resulting in multiple biases in recommendations, such as popularity bias \cite{Bias}. Moreover, the feedback loop in recommendation services leads to the ``Matthew effect,'' where the rich get richer \cite{zipf}. 
For traditional recommendation tasks, the previous efforts typically utilize additional regularization \cite{liu2023popularity,zhu2021popularity}, causal inference \cite{Causer,zhang2021causal}, and auxiliary knowledge \cite{text} to address this issue.

Regarding the specific bundle-aware tasks, PopCon \cite{PopCon} designs a popularity-based negative sampling strategy to mitigate its inherent popularity bias.
CoHeat \cite{CoHeat} leverages a popularity-based coalescence approach to address the highly skewed distribution of user-bundle interactions.
Notably, these works address popularity bias in user-bundle interactions by treating bundles as items, similar to traditional cold-start recommendation tasks.
However, DieT emphasizes the popularity of each item, achieving the promotion of long-tail items through the in-depth extraction of popularity-free knowledge and distilled knowledge transfer.


%% file: 3_motivation_analysis.tex
\section{Preliminary and Empirical Study}
\label{sec.ma}
We first introduce the product bundling task, with the specific definition of the Pop-to-LT scenario. Then, we conduct empirical studies to indicate the issues with existing methods and highlight the motivation for this work, including performance analysis of each scenario and surgical analysis of multiple modalities.

\subsection{Problem Formulation}
We aim to maximally capture the popularity-free knowledge and effectively transfer it into the bundling process to enhance the Pop-to-LT product bundling scenario.
To achieve this, we first define the set of items as $\mathcal{I}=\{i_1, i_2, \cdots, i_n\}$ with the corresponding textual features $\mathcal{T}=\{\bm{t}_1, \bm{t}_2, \cdots, \bm{t}_n\}$ and media feature $\mathcal{M}=\{\bm{m}_1, \bm{m}_2, \cdots, \bm{m}_n\}$, where $n$ denotes the quantity of items. 
The pre-extracted textual features are obtained from the relevant descriptions, titles, and attributes, and the media features are extracted from the visual or audio information. 
Additionally, we collect the interaction data between users and items to form the user-item interaction matrix $D=\{d_{ui}|u\in \mathcal{U}, i \in \mathcal{I}\}$, where $\mathcal{U}=\{u_1,u_2,...u_m\}$ denotes the set of users and $m$ is the number of users. Here, $d_{ui}=1$ indicates the presence of an interaction between user $u$ and item $i$, while $d_{ui}=0$ denotes absence.
For the product bundling task, we define the bundle $b=\{i_1,i_2,\cdots,i_{\widetilde{n}}\}$ as a set of several relevant items, where $\widetilde{n}=|b|$ denotes the number of items in bundle $b$.
Given a partial bundle $b_x \in b$, where $b_x$ is a subset of $b$ and adheres $|b_x|<|b|$, the goal of this task is to predict the missing item $i \in b\!\setminus\! b_x$ to complete the original bundle $b$. We define the specific setting where all items in $b_x$ are high-popularity, and items in $b\!\setminus\! b_x$ are long-tail as the Pop-to-LT product bundling scenario.

\begin{figure}[t]
    \centering 
    \includegraphics[width=1\linewidth]{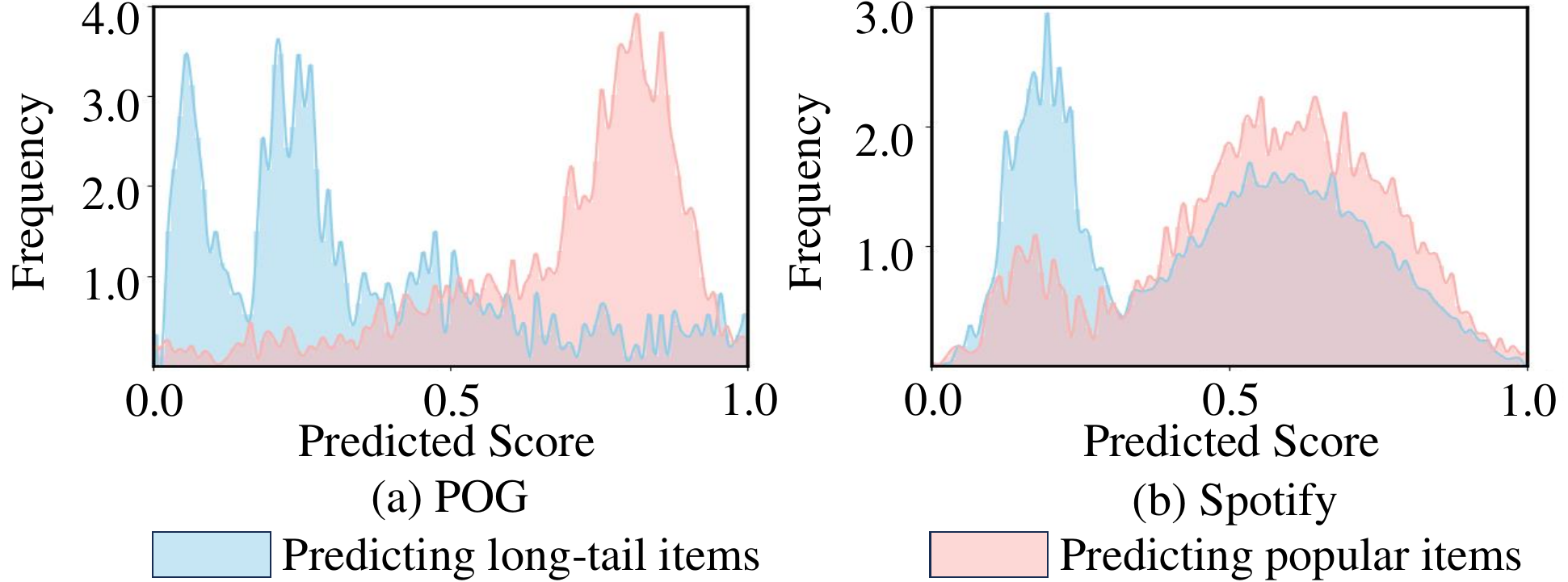} \\
    \caption{The score distribution for predicting long-tail and popular items of CLHE on two datasets.}
    \label{fig.performance}
\end{figure}

\subsection{Performance Analysis of Each Scenario}
\label{sec.ma_pa}
Due to differences in category, price, and brand, items in the corpus show diverse levels of popularity, with the presence of the “rich get richer” Matthew effect further intensifying this imbalance.
However, existing product bundling methods overlook this issue, thereby unfairly treating items with varying popularity.
It is intuitive that \textbf{\emph{they significantly suffer from the long-tail problem, resulting in poor accuracy for items with limited interactions.}}
To verify this opinion, we visualize the predicted probability distribution of the SOTA product bundling method \cite{CLHE} when predicting long-tail items and popular items on POG and Spotify. 
Figure ~\ref{fig.performance} indicates that the accuracy of CLHE is significantly lower when predicting long-tail items compared to popular ones. 
Notably, CLHE shows the most significant performance differences between different popularity settings in the fashion e-commerce dataset POG compared to the music-sharing dataset Spotify, which focuses more on item popularity. This phenomenon aligns perfectly with the intuitive assumptions of the Pop-to-LT product bundling scenario. In practical applications, this is a challenging and worthwhile issue to address.

\subsection{Surgical Analysis of Multiple Modalities}
\label{sec.ma_sa}

\begin{figure}[t]
    \centering
    \includegraphics[width=1\linewidth]{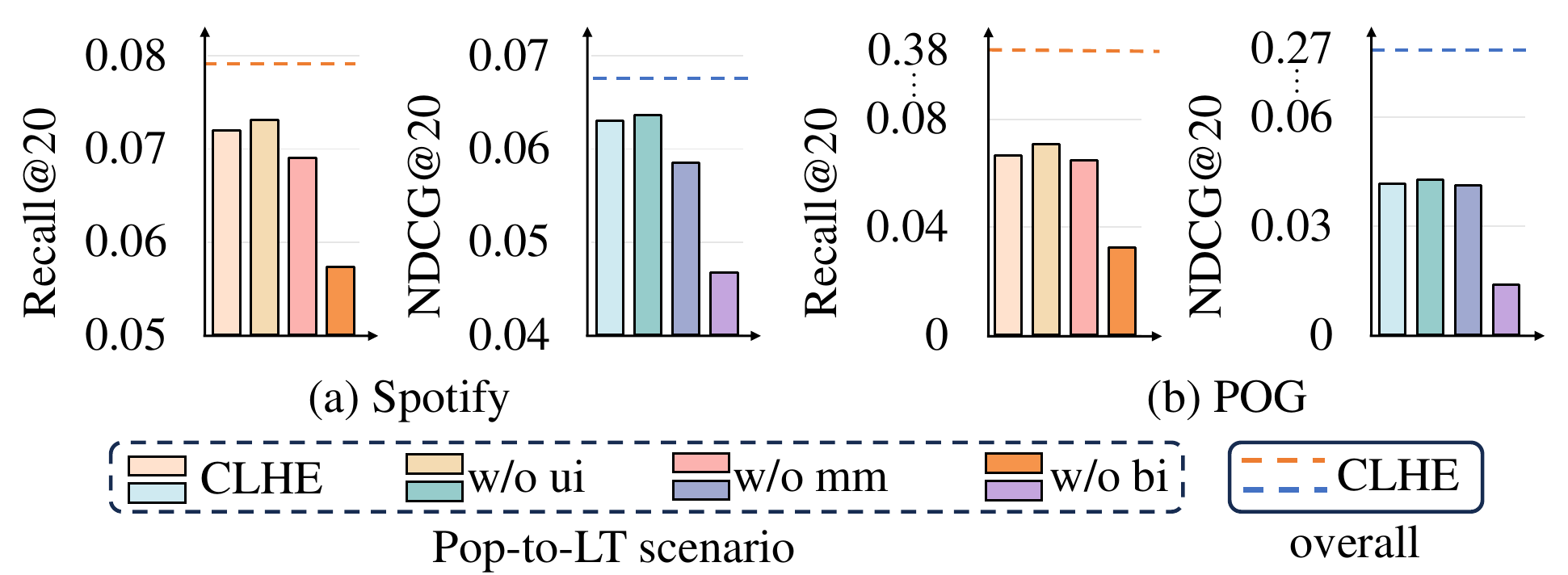}
    \caption{Surgical results of multiple modalities in CLHE on Pop-to-LT scenario. Removing bundle-item affiliation or multimodal features results in a significant drop.}
    \label{fig.h2l}
\end{figure}

As mentioned in Section.~\ref{sec.ma_pa}, existing bundling strategies exhibit diverse accuracy when predicting items with different levels of popularity.
In that case, it may encounter a performance collapse in the more challenging Pop-to-LT product bundling scenarios.
To verify this point, we compare CLHE \cite{CLHE} on this specific scenario with the overall product bundling scenario in Figure ~\ref{fig.h2l}. It reveals that:
In contrast to the full evaluation, \textbf{\emph{CLHE exhibits worse performance in the Pop-to-LT setting}}. This is due to its lack of explicit consideration of item popularity knowledge, which hinders its ability to achieve optimal performance in this more challenging and non-doctrinaire setting.
Jointly considering these discoveries with the specific definition of this scenario, we have intuitively sparked two bold questions: Are the pre-extracted item-level user feedback features responsible for the performance decline of current bundling methods in this scenario? 
Is the effectiveness of various modality-aware item representations inconsistent in the Pop-to-LT scenario?
To answer these questions, we conduct a surgical analysis to explore the effectiveness of each view in the Pop-to-LT scenario. From Figure ~\ref{fig.h2l}, we can observe that:
(1) CLHE w/o ui outperforms CLHE in the Pop-to-LT scenario, demonstrating the \textbf{\emph{low quality of item-level user feedback in Pop-to-LT scenario}}. 
Intuitively, the incorporation of the unbalanced user feedback knowledge may force the bundle constructor to predict the popular items, thereby aggravating the inherent popularity bias in bundle construction.
(2) Comparing CLHE, CLHE w/o mm, and CLHE w/o bi, we notice that the \textbf{\emph{item representations from the bundle-item view and multimodal view are indispensable in Pop-to-LT product bundling}}.
To sum up, all these findings collectively constitute the motivation of our work: Effectively utilizing task-related bundle-item affiliation information and popularity-irrelevant multimodal knowledge to counteract the popularity bias caused by the low-quality user-item information in existing product bundling methods.

%% file: 4_methodology.tex
\section{Methodology}
\begin{figure*}[t]
    \centering 
    \includegraphics[width=1\linewidth]{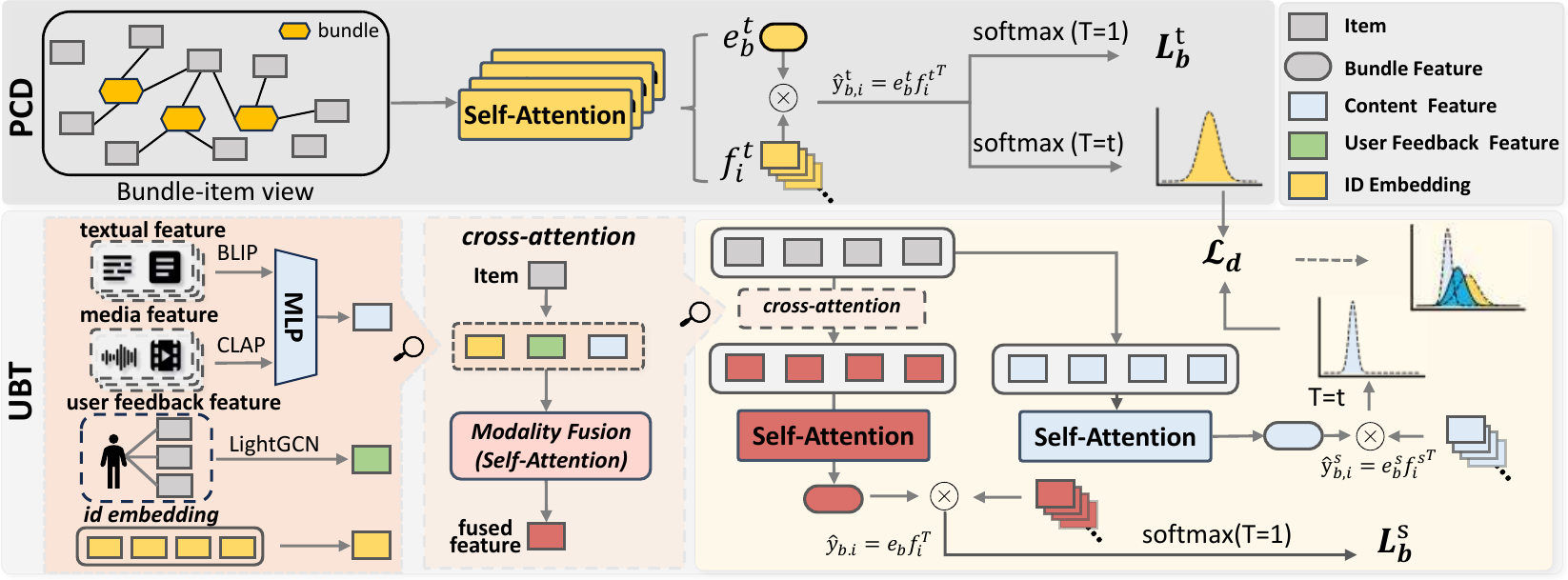} \\
    \caption{The overall structure of Diet. The PCD extracts the popularity-free bundle-item affiliation to guide the UBT for long-tail item promotion via a tailored knowledge distillation paradigm.}
    \label{fig.DieT}
\end{figure*}
To introduce the technical details of our Distilled Modality-Oriented Knowledge Transfer (DieT) framework, we first outline the product bundling backbone to illustrate the implementation process of this task. Inspired by the discoveries in Section.~\ref{sec.ma}, we provide in-depth definitions of the Popularity-free Collaborative Distribution Modeling (PCD) and Unbiased Bundle-aware Knowledge Transferring (UBT).
Finally, we detail the optimization and discussion of DieT.
\subsection{Product Bundling Backbone}
\label{sec.backbone}
Without loss of generality, DieT can be employed by any existing multimodal product bundling model. Inspired by the powerful hierarchical encoding capabilities on both the feature level and the item level of CLHE \cite{CLHE}, we adopt it as the backbone of our DieT.
Specifically, we first extract the textual feature $\bm{t}_i$ \cite{blip} and the media feature $\bm{m}_i$ \cite{clap} for item $i$ and then generate the fused content feature $\bm{c}_i = \text{average}(\bm{t}_i, \bm{m}_i)$. 
We also initialize the self-learnable item embedding $\bm{v}_i$ to capture item-bundle-level affiliation patterns and employ LightGCN \cite{lightgcn} to obtain the item-level user feedback $\bm{p}_i$.
Then, we concatenate these features from the same item $i$ to obtain the final item representation $ \bm{F}_i = \text{concat}(\bm{W}_c\bm {c}_i, \bm{W}_p\bm{p}_i, \bm{v}_i), \bm {F}_i \in \mathbb{R}^{3\times d}$. 
Here, $\text{concat}(\cdot)$ denotes the concatenation function, $\bm{W}_p$ and $\bm{W}_c$ represent the corresponding transformation matrices respectively. After that, a multi-layer self-attention encoder is utilized to model the correlations among these aforementioned features:
\begin{equation}
\begin{cases}
\bm {A}_i^{(l)}=\frac {1}{\sqrt{d}}\bm {\widehat{H}_i}^{(l-1)}\bm{W}_I^K(\bm {\widehat{H}_i}^{(l-1)}\bm{W}_I^Q)^{\intercal},\\
\bm {\widetilde{H}_i}^{(l)}= \text{softmax}(\bm{A}_i^{(l)})\bm {\widehat{H}_i}^{l-1},\\
\end{cases}
\label{eq.self-atten}
\end{equation}
where $\bm{W}_I^K \in \mathbb{R}^{d\times d}$ and $\bm{W}_I^Q \in \mathbb{R}^{d\times d}$ denote the linear projections which transpose the final item representation $\bm{F}_i$ into key and value spaces, $\bm {\widehat{H}_i}^{(0)} = \bm{F}_i$ and $\bm {\widetilde{H}_i}^{L}$ denotes the medium feature of item $i$ after $L$ layers of self-attention operator. Then we can generate the item's representation $\bm{f}_i \in \mathbb{R}^d $ by averaging the multiple features:
\begin{equation}
\bm{f}_i = \text{average}(\bm{\widetilde{H}_i}^{(L)}).    
\end{equation}

\par After obtaining the item representation $\bm{f}_i$, we build another bundle-level self-attention module to capture the representation $\bm{e}_b$ of the given partial bundle $b_x$. Similar to Eq (\ref{eq.self-atten}), we use a $Z$-layer self-attention encoder to encode the concatenation of item representations $\bm {f}_i$ in the partial bundle $b_x$. Two trainable matrices $\bm{W}_B^K \in \mathbb{R}^{d\times d}$ and $\bm{W}_B^Q \in \mathbb{R}^{d\times d}$ project these item embeddings into the key and value spaces. After that, we can obtain the final representation of the bundle $\bm{e}_b = \text{average}(\bm{\widetilde{E}_b}^{(Z)})$ by averaging the representations from each layer, where $\bm {\widetilde{E}_b}^{Z}$ denotes features' representation after $Z$ layers of self-attention operator.

\subsection{Popularity-free Collaborative Distribution Modeling}
From the previous empirical analyses in Section.~\ref{sec.ma}, we have noticed that the knowledge from different perspectives has varying effects in the Pop-to-LT scenario, and bundle-item affiliation knowledge plays a crucial role in this setting.
We have also pointed out that a vast number of items exhibit diverse popularity distribution in the user-item view (infrequent) and the bundle-item view (frequent), such as niche fashion outfit generation or music collections. 
For these items that frequently occur within bundles, the dynamic optimization of their representation during the training process can precisely capture the inherent bundle-item level affiliation knowledge for these long-tail items in Pop-to-LT scenarios. 
To achieve this, we propose a Popularity-free Collaborative Distribution Modeling module (PCD) to highlight the popularity-independent knowledge via the collaborative distribution in the specific bundle-item view.

Specifically, we first initialize the final item representation $\bm{f}_i^t$\footnote{We use superscript $t$ to denote variables in PCD and $s$ in UBT for differentiation.} of item $i$ with the bundle-level item embedding $\bm{v}_i$.
Then, we model the bundle representation $\bm{e}_b^t$ of the given partial bundle $b_x$ via the following Eq.~\ref{eq.self-atten_b} to capture the bundle-item affiliation knowledge:
\begin{equation}
\begin{cases}
\bm {A}_b^{(z)}=\frac {1}{\sqrt{d}}\bm {\widehat{E}_b}^{(z-1)}\bm{W}_B^K(\bm {\widehat{E}_b}^{(z-1)}\bm{W}_B^Q)^{\intercal},\\
\bm {\widetilde{E}_b}^{(z)}= \text{softmax}(\bm{A}_b^{(z)})\bm {\widehat{E}_b}^{z-1},\\
\end{cases}
\label{eq.self-atten_b}
\end{equation}
where $\bm{\widehat{E}_b}^{(0)}\!=\!\text{concat}(\{\bm {f}_i^t\}_{i \in b})$. Subsequently, we average the multiple features to obtain the bundle representation $\bm{e}_b^t\!=\!\text{average}(\bm{\widetilde{E}_b}^{(Z)})$. 
After these three steps, we can compute the matching scores, which indicates the likelihood of item $i$ being included in the partial bundle $b_x$ to form a complete bundle via the inner product as:
\begin{equation}
    \label{eq.scores}
    \widehat{L}ogits_{b,i}^t = \bm{e}_{{b}}^t{(\bm{f}_{i}^t)}^{\intercal}.
\end{equation}

While relying solely on bundle-item level features makes it struggle to outperform existing product bundling methods (see Section.~\ref{sec.ma_sa}), it effectively retains the bundle-item affiliation information through this simple yet universal design, thereby supporting the bundle representation modeling in transferring.

\subsection{Unbiased Bundle-aware Knowledge Transferring}
Intuitively, deliberately omitting the information of bundles from any view may lead to a defective bundle representation modeling process, thereby resulting in decreased product bundling performance. In contrast, the empirical analyses employed in Section.~\ref{sec.ma_sa} reveal that certain views may mislead the optimization process in the specific scenario of Pop-to-LT product bundling. This prompts us to consider how to efficiently utilize the popularity-free collaborative knowledge to boost Pop-to-LT product bundling tasks without excessively impacting overall product bundling. Thus, we propose an Unbiased Bundle-aware knowledge Transferring module (UBT) to solve this issue.

Specifically, we first leverage a cross-modality mapping layer to generate the content feature $\bm{c}_i^s$ for item $i$ using the pre-extracted textual feature $\bm{t}_i$ and media features $\bm{m}_i$, which is formulated as:
\begin{equation}
\bm{c}_i^s=\text{average}(f_{\mathcal{T}}(\bm{t}_i), {f}_{\mathcal{M}}(\bm{m}_i)),
\end{equation}
where the mapping functions $f_{\mathcal{T}}(\cdot)$ and $f_{\mathcal{M}}(\cdot)$ can provide subtle modifications for both textual and multimedia features to enable them to accommodate the product bundling task. 
Subsequently, we utilize the self-attention modality fusion method to generate the final item representation $\bm{f}_i^s$ with the content features $\bm{c}_i^s$, the item-level user features $\bm{p}_i$, and the bundle-item affiliation features $\bm{v}_i$.
Then, we leverage the same bundle-level self-attention function in Section.~\ref{sec.backbone} to model the bundle representation $\bm{e}_b^s$ of the partial bundle $b_x$. These two features are utilized to optimize the product bundling task as the primary construction loss in Eq (\ref{eq.main_loss}).

With the discoveries in Section.~\ref{sec.ma_sa}, we further rectify the content knowledge in UBT to zoom in on the Pop-to-LT product bundling task.
Specifically, the content knowledge of each item is entirely popularity-free, as it only deals with multimodal information, making it entirely unrelated to the item's popularity in the user-item view. In this case, we follow the knowledge distillation paradigm to transfer the bundling-related parts of collaborative knowledge in the bundle-item view into the content features, thereby obtaining more robust and accurate content representations for Pop-to-LT scenarios.
That is, we utilize content information to construct the item representation $\bm{f}_{i_c}^s = \bm{c}_i^s$ and modality-specific self-attention function to aggregate bundle representation $\bm{e}_{b_c}^s$. As in Eq (\ref{eq.scores}), we calculate the matching scores of UBT as:


%


\begin{equation}
\widehat {L}ogits_{b,i}^s = \bm{e}_{{b}_c}^s{(\bm{f}_{i_c}^s)}^{\intercal}.
\label{eq.logits}
\end{equation}
It is noteworthy that when calculating the scores in UBT, we employ the modality-specific bundle representation $\bm{e}_{{b}_c}^s$, to avoid UBT simply replicating the results of PCD, which would degrade overall performance.
More importantly, the significance of popularity-free features is further enhanced, thereby mitigating the negative impact of biased item-level user feedback features. 
After these steps, the crucial question is: How do we design the knowledge transfer method to assimilate the well-optimized bundle-item knowledge from PCD? 
We convert items' matching scores in Eq (\ref{eq.scores}, \ref{eq.logits}) into the corresponding probability distribution as follows:
\begin{equation}
\begin{cases}
    P_t=\text{softmax}(\frac {\widehat {L}ogits_{b,i}^t}{T}),\\
    P_s= \text{log\_softmax}(\frac{\widehat {L}ogits_{b,i}^s}{T}).
\end{cases} 
\end{equation}
Notably, we do not enforce UBT to match the probability distribution of PCD exactly. On the contrary, we desire UBT to learn the popularity-free and bundle-familiar knowledge from PCD via knowledge distillation. Therefore, we employ soft targets \cite{2015distilling} to achieve this primary goal. The distillation loss can be described as:
\begin{equation}
\mathcal{L}_d = \frac{1}{|\mathcal{B}|}\text{KL}(P_s||P_t) \times T^2,
\end{equation}
where $|\mathcal{B}|$ represents the batch size, $\text{KL}(\cdot)$ refers to the Kullback-Leibler (KL) divergence \cite{KL} between the probability distributions $P_s$ and $P_t$ of the PCD and UBT. 
Here, the temperature parameter $T^2$ is introduced in the loss computation process to ensure the appropriate scale of the distillation loss. We also consider feature-based distillation loss to replace the distribution-based one, thereby transferring the feature-level bundle-item affiliation knowledge from PCD to UBT, this distilled loss can be formulated as:
\begin{equation}
\mathcal{L}_d = \frac{1}{|\mathcal{N}|} \sum_{i=0}^{\mathcal{N}-1}\mathcal{L}(\text{sim}(\bm{e}_{{b}}^t,\bm{e}_{{b}_c}^s)),  
\end{equation}
where $\text{sim}(\cdot)$ represents the general similarity function (e.g., cosine, Euclidean, KL divergence, dot product, etc.). The primary focus of DieT is the logits distillation, while the effectiveness of feature-based distillation is verified in Section. \ref{Sec.valid}.

\vspace{-0.3cm}
\subsection{Optimization}
\label{sec.loss}
Given the item representation $\bm{f}_i$ of item $i$ and the bundle representation $\bm{e}_b$ of partial bundle $b_x$, we compute the corresponding matching score $\hat{y}_{{b},i}={\bm{e}_b}\bm{{f}_i}^{\intercal}$ via the inner-product function.
Following the classical bundle construction work \cite{CLHE}, we leverage the negative log-likelihood loss 
to optimize the product bundling task:
\begin{equation}
\label{eq.main_loss}
\mathcal{L}_b=-\frac {1}{|\mathcal{B}||\mathcal{I}|} \sum _ {b\in \mathcal{B}} \sum_{i \in \mathcal{I}} {y}_{\bm{b},i} \text{log\_softmax}_b( \hat{y}_{{b},i}).
\end{equation}


Then, the overall objective function $\mathcal{L}$ can be formulated as a linear combination of the construction loss $\mathcal{L}_{b}$, the knowledge distillation loss $\mathcal{L}_d$ and the regularization term as:
\begin{equation}
\mathcal{L}=\mathcal{L}_{b}+\lambda\mathcal{L}_d + \beta ||\bm{\theta}||_ {2}^ {2},
\end{equation}
where $\lambda$ denotes the weight of the knowledge distillation loss $\mathcal{L}_d$. The term $||\bm{\theta}||_ {2}^ {2}$ represents the $L_2$ regularization term, where $||\bm{\theta}||$ denotes all the trainable parameters in our DieT.
\label{sec.kd_method}

\vspace{-0.1cm}
\subsection{Model Discussion}
CLHE \cite{CLHE} is a pioneering work in the field of product bundling and achieves the SOTA performance.
However, its accuracy is significantly lower when predicting long-tail items compared to popular ones (\ie the Pop-to-LT scenario), which may be attributed to the mechanical multi-view knowledge fusion. In contrast, DieT proposes a knowledge distillation framework to more effectively utilize bundle-item affiliation information and popularity-independent content information to mitigate the low-quality user-item information in existing bundles, thereby reducing popularity bias. 
Notably, DieT achieves significant performance improvements in Pop-to-LT scenario while also delivering remarkable gains in overall bundling scenario with the meticulous design of these two modules.

%% file: 5_experiments.tex
\section{Experiments}
We conduct extensive experiments on two real-world datasets to answer the following research questions:
\begin{itemize}[leftmargin=*, topsep=0.2pt,parsep=0pt]
\item \textbf{RQ1:} How does DieT perform against the SOTA methods on the full product bundling task and the Pop-to-LT bundling scenario?
\item \textbf{RQ2:} Is DieT still effective and robust with diverse distillation strategies, popularity ratio, and bundling scenarios?
\item \textbf{RQ3:} Does DieT demonstrate the generalization with other base product bundling backbones?
\item \textbf{RQ4:} How does DieT contribute to the Pop-to-LT product bundling compared to existing SOTA bundling methods? 
\end{itemize}


\begin{table}[!t]
    \caption{Statistics of two real-world bundling datasets.}
    \vspace{-0.3cm}
    \centering
    \fontsize{9pt}{9pt}\selectfont
    \begin{tabular}{lccccccccc}
        \toprule
        Dataset& \#U& \#I& \#B& \#B-I& \#U-I\\
        \midrule
        Spotify& 118,899& 72,453&12,309 &359,935 & 21,032,433\\
        POG&  2,311,431& 31,217&  29,686&  105,775& 6,345,137\\
        \bottomrule
    \end{tabular}   
    \label{dataset}
\end{table}

\subsection{Experimental Settings}
\subsubsection{Datasets}
The application scenarios for product bundling are diverse, including e-commerce, meal, gaming platforms, \etc \cite{pog, Spotify}. 
To evaluate our method, we select two datasets from the fashion domain POG and the music playlist dataset Spotify because they encompass the three types of features mentioned earlier.
Additionally, to evaluate the effectiveness of our method in Pop-to-LT scenarios, we utilize user-item interaction data to measure popularity. 
Items are sorted by popularity, and we select a certain proportion of high-popularity items from the head and low-popularity items from the tail. 
Specifically, while keeping the items within the bundle unchanged, we only modify whether they act as inputs or targets in the test set. 
For each dataset, we randomly divide all bundles into training, test, and validation sets in a 7:2:1 ratio. The details of the datasets are presented in Table. \ref{dataset}.

\subsubsection{Baselines}
Considering product bundling has been less explored yet, we observe a scarcity of research that precisely matches our framework.
Therefore, we select several leading methods and tailor them to our specific settings. For a fair comparison, all baseline methods utilize three types of features as our method and employ the identical negative log-likelihood loss.
Additionally, we incorporate debiasing methods from traditional recommendation tasks into the best baseline and compare them with our method.
\\ \textbf{General Product Bundling methods:}
\begin{itemize}[leftmargin=*, topsep=0.2pt,parsep=0pt]
\item \textbf{MultiDAE} \cite{MultiDAE} adopts an auto-encoder architecture and utilizes average pooling to aggregate item representations.
\item \textbf{MultiVAE} \cite{MultiVAE} is a variational auto-encoder model built upon MultiDAE, incorporating variational inference techniques.
\item \textbf{Bi-LSTM} \cite{Bi-LSTM} treats each bundle as a sequence and employs bidirectional LSTM to learn the bundle representation.
\item \textbf{Hypergraph} \cite{Hypergraph} represents each bundle as a hyper-graph and utilizes a GCN model to learn the bundle representation.
\item \textbf{Transformer} \cite{wei2023strategy} customizes a transformer to capture item interactions and generate the bundle representation.
\item \textbf{CLHE} \cite{CLHE} is a contrastive learning-enhanced hierarchical encoder method that achieves SOTA performance. 
\end{itemize}
\textbf{Debiased Product Bundling methods:}
\begin{itemize}[leftmargin=*, topsep=0.2pt,parsep=0pt]
\item \textbf{IID} \cite{iid} achieves unbiased recommendations by training item embeddings that maintain popularity neutrality in all directions.
\item \textbf{PID} \cite{neutralizing} works by isolating a popularity-related direction and correcting it post-training to neutralize popularity bias.  
\end{itemize}

\subsubsection{Implement Details}
Following the pioneering work \cite{CLHE}, we define the embedding size and the dimension of hidden representations as 64. We use Xavier initialization \cite{Xavier}, Adam optimizer \cite{Adam}, and batch size 2048 for DieT and the compared baselines. We embark on a grid search to identify the optimal hyper-parameter settings. Specifically, we explore a range of learning rates within $\{10^{-2}, 10^{-3}, 5\times10^{-3}, 10^{-4}, 5\times10^{-4}\}$, and adjust $\beta$ within the range of $\{10^{-3}, 10^{-4}, 10^{-5}, 10^{-6}\}$. In most cases, optimal performance is achieved when $\beta$ is set to $10^{-5}$. In POG, the optimal learning rate is $10^{-3}$, while in Spotify, it is $10^{-4}$. 
In knowledge distillation, we explore the temperature $T$ within the range of $\{1, 2, 3\}$ and the parameter $\lambda$ for distillation loss within the range of $\{0.5, 1.0, 1.5, 2.0\}$. In most cases, optimal performance is achieved with $T=2$ and $\lambda=1.0$. We follow the design described in their paper for the baseline model to achieve optimal performance.


\begin{table*}[!t]
\vspace{-0.2cm}
\caption{Performance comparison between DieT and baselines on Pop-to-LT product bundling scenario on two datasets.}
\vspace{-0.2cm}
\begin{tabular}{l|lcccccc|cc|cc}
\toprule
Dataset & Metric & MultiDAE & MultiVAE & Bi-LSTM & Hypergraph & Tranformer & CLHE & IID & PID & \textbf{DieT} & \textbf{\#Improv.} \\ \midrule
\multirow{4}{*}{Spotify} & R@20 & 0.0172   & 0.0121   & 0.0385  & 0.0614   & 0.0166    & {0.0741}    &0.0737    &\underline{0.0745} & \textbf{0.0860}  & \textbf {15.43\%}     \\
\multicolumn{1}{c|}{} & N@20 & 0.0125   & 0.0086   & 0.0320 & 0.0539    & 0.0119    & {0.0645}    &\underline {0.0649}     &\underline {0.0649} & \textbf{0.0772} & \textbf{18.95\%}     \\
\multicolumn{1}{c|}{} & R@40 & 0.0339  & 0.0252  & 0.0621  & 0.1000    & 0.0319     & \underline {0.1168}  &0.1136 &0.1159    & \textbf{0.1283} & \textbf{9.85\%}      \\
\multicolumn{1}{c|}{} & N@40 & 0.0199   & 0.0144   & 0.0426  & 0.0713    & 0.0187     & \underline {0.0837}   &0.0828    &0.0835   & \textbf{0.0959} & \textbf{14.58\%}     \\ \midrule

\multirow{4}{*}{POG} & R@20 & 0.0377   & 0.0570    & 0.0121  & 0.0551     & 0.0460     &  {0.0661}     &\underline{0.0707}& 0.0691 & \textbf{0.0773} & \textbf{9.34\%}     \\
\multicolumn{1}{c|}{}  & N@20 & 0.0160    & 0.0285   & 0.0055  & 0.0238     & 0.0262     &  {0.0392}     &0.0417&\underline{0.0419}  & \textbf{0.0430}  & \textbf{2.62\%}      \\
\multicolumn{1}{c|}{} & R@40 & 0.0508   & 0.0755   & 0.0228  & 0.0725     & 0.0636    &  {0.0904}     &0.0911& \underline{0.0925}  & \textbf{0.0989} & \textbf{6.91\%}      \\
\multicolumn{1}{c|}{} & N@40 & 0.0189   & 0.0326   & 0.0078  & 0.0277     & 0.0301    & {0.0446}   &\underline{0.0470}&0.0467    & \textbf{0.0479} & \textbf{1.91\%} \\ \bottomrule
\end{tabular}
\label{table.hl}
\end{table*}
\begin{table*}[!t]
\caption{Performance comparison between DieT and baselines on full product bundling scenario on two datasets.}
\vspace{-0.2cm}
\begin{tabular}{l|lcccccc|cc|cc}
\toprule
Dataset & Metric & MultiDAE & MultiVAE & Bi-LSTM & Hypergraph & Tranformer & CLHE   & IID & PID & \textbf{DieT}& \textbf{\#Improv.} \\ \midrule

\multirow{4}{*}{Spotify} & R@20 & 0.0614   & 0.0534   & 0.0411  & 0.0733     & 0.0672     & \underline {0.0815} &0.0812 & 0.0796& \textbf{0.0832} & \textbf{2.09\%}      \\
\multicolumn{1}{l|}{} & N@20 & 0.0541   & 0.0470    & 0.0348  & 0.0619     & 0.0590      &  {0.0694} &\underline{0.0697} & 0.0673 & \textbf{0.0711} & \textbf{2.01\%}      \\
\multicolumn{1}{l|}{} & R@40 & 0.0971   & 0.0836   & 0.0675  & 0.1149     & 0.1078     &  {0.1286} &\underline{0.1290} & 0.1253 & \textbf{0.1322} & \textbf{2.48\%}      \\
\multicolumn{1}{l|}{} & N@40 & 0.0700     & 0.0605   & 0.0467  & 0.0806     & 0.0772     &  {0.0905} &\underline{0.0911} & 0.0878 & \textbf{0.0930}  & \textbf{2.09\%}      \\ \midrule
\multirow{4}{*}{POG}  & R@20 & 0.3082   & 0.3076   & 0.3141  & 0.2695     & 0.3035     &  {0.3811} &\underline{0.3820} &0.3815 & \textbf{0.3842} & \textbf{0.11\%}      \\
\multicolumn{1}{l|}{} & N@20 & 0.2008   & 0.1891   & 0.1972  & 0.1815     & 0.1846     & \underline {0.2776} &0.2704 &0.2754 & \textbf{0.2779} & \textbf{0.11\%}      \\
\multicolumn{1}{l|}{} & R@40 & 0.3367   & 0.3346   & 0.3423  & 0.2888     & 0.3244     &  {0.4090} &\underline{0.4092} &0.4086  & \textbf{0.4113} & \textbf{0.51\%}      \\
\multicolumn{1}{l|}{} & N@40 & 0.2074   & 0.1954   & 0.2039  & 0.1861     & 0.1895     & \underline {0.2842} &0.2769 &0.2818 & \textbf{0.2843} & \textbf{0.04\%} \\ \bottomrule
\end{tabular}
\label{table.overall}
\end{table*}

\subsection{Performance Comparison (RQ1)}
To demonstrate DieT's effectiveness, we conduct tests from both Pop-to-LT and full product bundling scenario perspectives, comparing them with multiple baselines on real-world datasets from two domains. As illustrated in Table. \ref{table.hl} and Table. \ref{table.overall}, we accentuate the best results on the same datasets with bold font and underline the best baselines. Popular ranking metrics such as Recall@K and NDCG@K are employed ~\footnote{For simplification, we leverage N@k and R@k to indicate the NDCG@k and Recall@k in the following figures and tables.}. Our observations are as follows: 
\begin{itemize}[leftmargin=*, topsep=0.2pt,parsep=0pt]
\item[1)] In the Pop-to-LT scenario, which is the main focus of our work, DieT beats all the baselines on all datasets, demonstrating the effectiveness of our popularity-free collaborative distribution modeling and unbiased bundle-aware knowledge transferring;
\item[2)] Improvement of DieT to the backbone correlates with the average number of items per bundle, indicating that the PCD, which excavates more comprehensive bundle-item affiliation information, is more effective in guiding the UBT. Hence, DieT shows notable improvement in both scenarios on Spotify;
\item[3)] DieT achieves comparable results to the optimal baseline model overall, attributed to UBT’s well-designed knowledge transfer mechanism, which leverages content features to integrate popularity-free collaborative knowledge from PCD.
\item[4)] The improvement of DieT at k=20 is greater than at k=40, indicating our method can more accurately identify the most relevant long-tail items in smaller recommendation lists.
\item[5)] We evaluate two viable debiasing methods from traditional recommendations. The findings indicate that these methods are not entirely suitable for product bundling tasks and are surpassed by DieT. Moreover, IID incurs substantial computational overhead, resulting in catastrophic runtime on large datasets.
\end{itemize}

\subsection{In-depth Analyses of DieT (RQ2)}
\subsubsection{Effectiveness Analysis of Diverse Distillation Methods.}\label{Sec.valid}
As we introduce in Section. ~\ref{sec.kd_method}, there are multiple distillation methods. Here, we employ different distillation methods within our DieT framework to test the effectiveness of each approach.
The results in Table. \ref{table.methods} demonstrate that regardless of the distillation method employed, UBT equipped with knowledge significantly improves in the Pop-to-LT scenario. Different methods achieve varying improvements. However, simply stacking these methods does not yield better results. The reason lies in their different optimization objectives and constraints, which lead to inconsistencies when used simultaneously, thus affecting the overall effectiveness. Therefore, to save computational resources, we opted to use Logits distillation. Here, we simply elucidate that our approach proves effective under various knowledge distillation methodologies. As for more detailed investigations, such as various loss calculations, additional optimization techniques, or how to better integrate the advantages of both distillation methods, we leave those for future work.
\begin{table}[!t]
\caption{Results of diverse knowledge distillation methods in DieT on the dataset of  POG. Here, FD stands for feature distillation, and LD stands for logits distillation.}
\fontsize{9pt}{9pt}\selectfont
\begin{tabular}{lcccc}
\toprule
Method         & R@20 & N@20 & R@40 & N@40 \\ \midrule
CLHE           & 0.0661   & 0.0392 & 0.0904   & 0.0446 \\ \midrule
DieT (LD) & \textbf{0.0773}   & 0.0430 & \textbf{0.0989}   & 0.0479 \\ 
DieT (FD) & 0.0705   & \textbf{0.0459} & 0.0950   & \textbf{0.0513} \\ 
DieT (FD+LD)  & 0.0730	&0.0416	&0.0893	&0.0452\\ \bottomrule
\end{tabular}
\label{table.methods}
\end{table}


\begin{figure}
    \vspace{-0.2cm}
    \centering
    \includegraphics[width=0.9\linewidth]{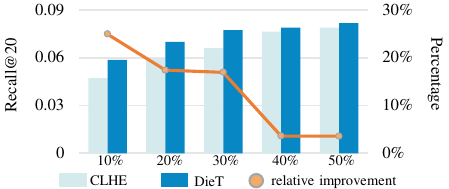}
    \caption{Performance comparison between DieT and SOTA method CLHE on diverse popularity ratios on POG.}  
    \label{fig.percent}
\end{figure}

\subsubsection{Robustness Analysis on Diverse Popularity Ratios}
In Table. \ref{table.hl}, taking into account the impact of popularity and the size of the test set, we categorize items based on their popularity rank into top-30\% popular and top-30\% long-tail items. It should be noted that as the definition of popularity becomes stringent, the number of bundles meeting the criteria decreases. 
Here, we select five different popularity intervals in the test set and compare the performance of the backbone with our model.
As the popularity requirements increase, Figure. \ref{fig.percent} clearly shows a noticeable performance drop in the backbone model, highlighting the challenge in more demanding scenarios, while the DieT model experiences a smaller decline and shows a certain improvement relative to the backbone models.
At the same time, we represent the improvement percentage with orange lines under different proportions. As the selection criteria become stricter, the improvement of our model becomes more apparent. This further underscores the knowledge transfer between the two modules of DieT in mitigating popularity bias issues. 

\begin{table}[!t]
\vspace{-0.1cm}
\caption{Performance comparison between DieT and SOTA method CLHE on diverse bundling scenarios on POG.}
\fontsize{9pt}{9pt}\selectfont
\begin{tabular}{llcccc}
\toprule
Scenario             & Model & R@20   & N@20   & R@40   & N@40   \\ \midrule
\multirow{2}{*}{overall} & CLHE       & 0.3811 & 0.2776 & 0.4090 & 0.2842 \\
                     & \textbf{DieT}       & \textbf{0.3842} & \textbf{0.2779} & \textbf{0.4113} & \textbf{0.2843} \\ \midrule
\multirow{2}{*}{LT-to-Pop} & CLHE       & \textbf{0.8571} & 0.6326 & 0.8763 & 0.6373 \\
                     & \textbf{DieT}       & 0.8566 & \textbf{0.6334} & \textbf{0.8832} & \textbf{0.6500} \\ \midrule
\multirow{2}{*}{Pop-to-Pop} & CLHE       & 0.4247 & 0.2949 & 0.4688 & 0.3046 \\
                     & \textbf{DieT}       & \textbf{0.4312} & \textbf{0.2962} & \textbf{0.4693} & \textbf{0.3049} \\ \midrule
\multirow{2}{*}{Pop-to-LT} & CLHE       & 0.0661 & 0.0392 & 0.0904 & 0.0446 \\
                     & \textbf{DieT}       & \textbf{0.0773} & \textbf{0.0430} & \textbf{0.0989} & \textbf{0.0479} \\ \bottomrule

\end{tabular}
\vspace{-0.5cm}
\label{tab.Scenario}
\end{table}

\subsubsection{Robustness Analysis on Diverse Bundling Scenarios}
Here, we intend to delve deeper into the performance of DieT across the various scenarios we previously analyzed.
As shown in Table. \ref{tab.Scenario}, we compare the results of DieT and CLHE across these four different scenarios. Firstly, it is evident that the model's performance naturally surpasses random overall results when predicting high-popularity items. This is particularly evident in the LT-to-Pop scenario. The results indicate that DieT has achieved significant improvements in the Pop-to-LT scenario while obtaining results comparable to CLHE in other scenarios. 
This proves that our carefully designed knowledge transfer strategy maintains the strength of the UBT and integrates the PCD's capability to handle long-tail items, making it a competitive solution across various application scenarios.




\begin{figure}
    \centering
    \includegraphics[width=1\linewidth]{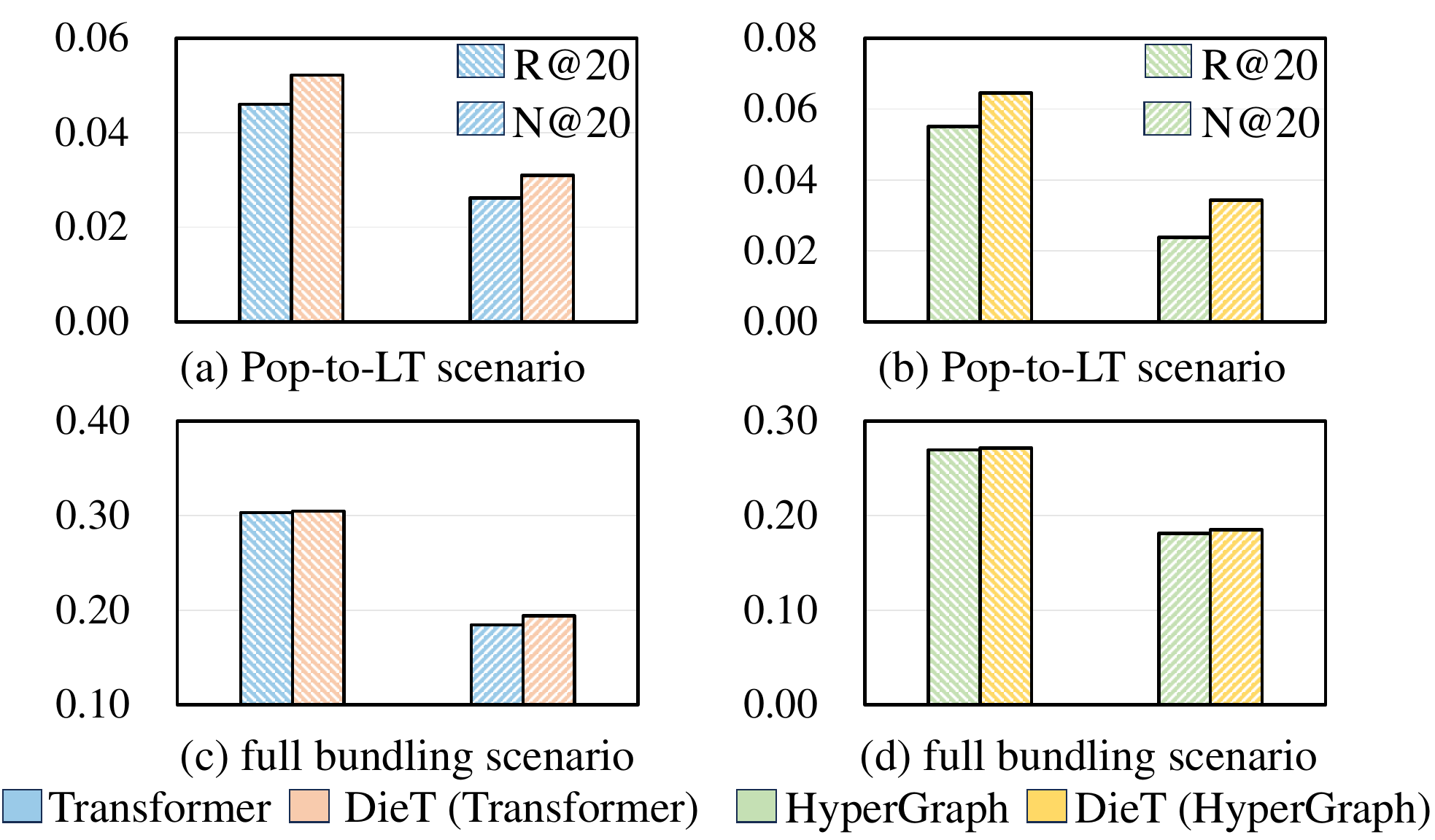}
    \caption{Generalization analysis of applying DieT to other backbones of Transformer and Hypergraph.}
    \label{fig.universal}
\end{figure}

\begin{figure*}[t]
    \centering
    \includegraphics[width=1\linewidth]{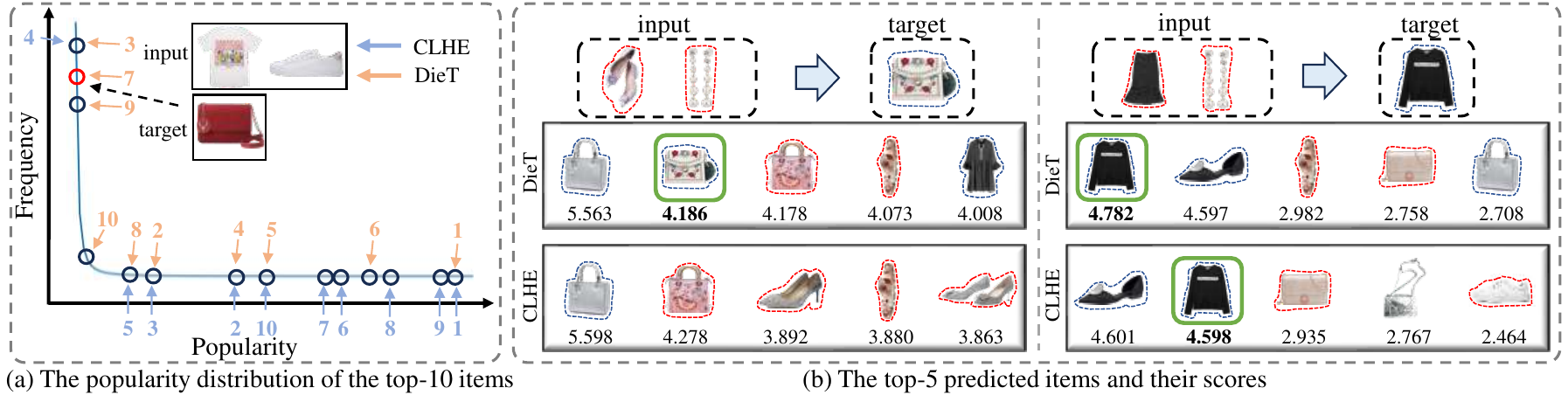}
    \caption{Case study of DieT. (a) The popularity distribution of top-10 items predicted by DieT and the baseline CLHE for several given popular items, with numbers representing the score ranking of the items. (b) The top-5 items and their scores predicted by DieT and CLHE for several given popular items, with red and blue dashed lines marking the \textcolor[RGB]{235,44,39}{popular} and \textcolor[RGB]{72,116,203}{long-tail} items.}
    \label{fig.case}
\end{figure*}

\subsection{Generalization Analysis of DieT (RQ3)}
To further demonstrate our approach's effectiveness, we modify several baseline methods that show promising results. We integrate our framework into the Transformer and HyperGraph methods. We keep the other parts of the baseline unchanged. Consistent with the previous description, we use a model with a self-attentive bundle-item affiliation excavator as the PCD, and these two models with three types of features as UBK's backbone. We compare the results of the models before and after incorporating DieT in terms of Pop-to-LT and full bundling scenario performance. The results in Figure. \ref{fig.universal} reveal that even in relatively simple models, our approach achieves significant improvements in the Pop-to-LT scenario and some enhancements in full bundling scenario performance. This indicates that DieT is a modality-oriented knowledge transfer framework independent of specific models, demonstrating the Generalization of our approach across different models.

\subsection{Case Study (RQ4)}
We would like to illustrate some cases to further describe how DieT improves performance in Pop-to-LT scenarios compared to CLHE. Specifically, we select the top 30\% items by popularity as popular items and the bottom 30\% as long-tail items and examine DieT's performance from two perspectives.
In Figure. \ref{fig.case} (a), we illustrate the distribution of the predictions on a general popularity distribution curve, where arrows of two colors represent the items predicted by each model.
Compared to CLHE, DieT not only correctly predicts target items within the top ten but also considers lower-popularity items more evenly, while CLHE tends to select high-popularity items for bundle construction (as indicated by the clustering of blue arrows towards the right side of the popularity spectrum). 
From another perspective, Figure. \ref{fig.case} (b) illustrates the score of the top five items between CLHE and DieT in Pop-to-LT scenarios. DieT achieves higher scores for low-popularity target items compared to CLHE. This is particularly evident in the left example, where DieT successfully predicts the target among the top five highest-scored items. Additionally, as we mark, DieT's construction results include more long-tail items, demonstrating its superior performance in promoting long-tail items in Pop-to-LT scenarios.
\FloatBarrier

%% file: 6_conclusion.tex
\section{Conclusion and Future Work}
In this work, we systematically investigated bundle construction issues to achieve combinations better aligned with real-world applications by addressing inherent popularity biases.
We analyzed existing challenges, highlighting uncertainties from user feedback data and suboptimal performance in Pop-to-LT scenarios. 
To address these, we proposed the Distilled Modality-Oriented Knowledge Transfer framework (DieT), which outperforms leading techniques on datasets from two domains. Comprehensive analysis and experimentation confirm DieT's effectiveness.

Despite the promising results, there is still considerable room for exploration in bundle construction. Firstly, 
is optimizing the extraction of user feedback features in an end-to-end manner possible?
Secondly, implementing popularity-based bundle generation strategies tailored to individual users is a promising direction for future research. Lastly, exploring how bundle construction techniques can be applied across different domains (\eg books and music, electronics and accessories) to create more diverse and appealing bundles is an interesting area for investigation.